# THE "POWER" DIMENSION IN A PROCESS OF EXCHANGE

Alberto Banterle

Contract Professor at University of Trieste

Department of Mathematical Economics and Statistics.

# Abstract

The field of study of this paper is the analysis of the exchange between two subjects. Circumscribed to the micro dimension, it is however expanded with respect to standard economic theory by introducing both the dimension of power and the motivation to exchange.

The basic reference is made by the reflections of those economists, preeminently John Kenneth Galbraith, who criticize the removal from the neoclassical economy of the "power" dimension. We have also referred to the criticism that Galbraith, among others, makes to the assumption of neoclassical economists that the "motivation" in exchanges is solely linked to the reward, to the money obtained in the exchange.

We have got around the problem of having a large number of types of power and also a large number of forms of motivation by directly taking into account the *effects* on the welfare of each subject, regardless of the means with which they are achieved: that is, referring to everything that happens in the negotiation process to the potential or real variations of the welfare function induced in each subject due to the exercise of the specific form of power, on a case by case basis, and of the intensity of the motivation to perform the exchange. In the construction of a mathematical model we paid great attention to its usability in field testing.

**Index:**

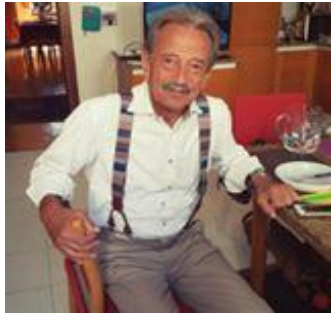

**Alberto Banterle** was born in Verona on 12-12-1940.

He attended school in Milan, obtaining a degree in Electrical Engineering at the Milan Polytechnic.

He began working in Milan in the telecommunications field, first within the Fiat Group, as Head of the Organization Department in a company of 9,000 employees, and then in the Alcatel Group until 1998.

Since 1999, he has been a contract professor at the University of Trieste for the course of Business Statistics in the Faculty of Economics, now the Department of Mathematical Economics and Statistics.

In these years he has been the coordinator of first level Masters degrees in e-Business and e-Government to which the Faculties of Engineering, Economics and Psychology have contributed. He has also promoted and managed regional projects for high schools on multimedia education topics.

**Foreword:**

The short essay that follows *does not claim to be a rigorous academic work*. It questions concepts such as "power", "motivation" and "welfare functions" behind which there is a vast literature in sociology, psychology and economics, which is largely unexplored by the author.

Nevertheless, in the hope of providing some meaningful suggestions, a formal elaboration of the mechanism that governs exchanges in simple situations that involve only two individuals is proposed below.

The market has been theorized to overcome the direct exchange between two people, but many of these exchanges are not realized through the market, but just between two subjects.

Attention is focused on cases in which there is an "*imbalance of power*" between the contracting parties. There are many such cases, particularly in those situations where authority prevails in its violent forms: blackmail and corruption.

A meaningful example is the assignment of a contract by a public body to a private company, without resorting - by necessity or political opportunity - to tenders or auctions to choose the best offer, or by distorting the mechanism to choose ex post the criteria that identify the desired ex ante supplier. In this case the two representatives of the public body and of the private company find themselves negotiating in a situation of potential abuse of power due to the absence of those constraints which, in a call for tenders, guarantee the identification of the most convenient choice for the public body.

The aim is to arrive at determining *objective* economic parameters (such as the price at the end of a negotiation) taking into account *subjective* perceptions concerning especially the imbalances of power between the parties involved.

Since exchange is the basic mechanism, present at all levels, which influences the way a society operates, it is hoped that this paper may contribute to the construction of models that might, for example, simulate the economic mechanisms of a society in which, for structural reasons or otherwise, power imbalances are widespread.

The work develops as follows: initially two determinants of the initial conditions for an exchange are identified and represented. These determinants are those assumed as exogenously given by the traditional economic analysis and they consist of the *motivational element*, i.e. the identification of the force of the reciprocal motivation to exchange and the *element of power*, which is the imbalance of strength between the two parties (paragraphs 1 and 2). Every initial unequal distribution in the motivation, and in the power, results in an alteration of the reserve prices, that is of the maximum willingness to pay and minimum to receive, even before the negotiation begins. In paragraph 3) a model is proposed for the representation of the negotiation process, which leads to the reaching of an agreement that is still affected by the imbalances of motivation and power. In paragraph 4) the model is extended from the isolated case to the long chains of exchanges that can be implemented across the planet. It is argued that the price set by the subject who enters the market is the result of the presence of imbalances stemming from the various stages in the supply chain.

Next, (paragraph 5) the same motivational and power imbalances are represented and analyzed with reference to *non-market exchanges*, i.e. exchanges whose outcome is not conditioned by the fixing of a price. In paragraph 6) the emphasis shifts to the *weak subjects* to identify the possible strategies with which they can defend themselves from the strong subjects. In paragraph 7) a reflection is made on the possibility of constructing models, passing from the micro to the macro, to represent the economic system within authoritarian and corrupt societies, where at all levels there are types of exchanges characterized by imbalances of motivation and power between the "strong" "and the "weak ".

## 1) Starting points.

Below is an analysis, in a simplified and brief form, of the elements that come into play in a process of exchange, of goods or services, between two subjects.

At the center of the examination we have, firstly, in terms of conceptual reference, ***the subjective welfare or utility function***[1] of those who are preparing to start an exchange. As a relational aspect, then, in the context of exchange, the "***power dimension***" will be introduced which, as we know intuitively, plays a decisive role in the agreements reached at the end of the process.

That said, in a normal exchange not encumbered by particular imbalances, what activates the exchange process is, for each subject, the *"**motivation**"*[2].

*In this context we define as "motivation" the difference between the increase in well-being connected to what each subject is preparing to <u>receive</u> and the loss connected to what he is about to <u>yield</u>:*

$$M = \Delta U \qquad\qquad\qquad 1)$$

Where **ΔU** is the balance, in terms of welfare, between the positive variation connected to the acquisition and the negative variation connected to the sale.

This is not a balance made of easily quantifiable amounts and algebraic differences: the **ΔU** is rather manifested, in the case in which investigations were carried out in the field, in the subjective ***belief of earning*** enough or a lot, or a little, depending on the case, even before a negotiation starts.

It is the element that pushes, in a more or less strong way, the activation of the exchange process. In principle, the process is triggered by the most motivated subject.

---

[1]
In this paper we do not refer to particular expressions of the utility function or even to subjective probabilities: we take into consideration very simple situations in which we make comparisons only in terms of convenience or perceived imbalances

[2]
The term "motivation" is widely used in psychology and in particular in human resource management studies. In this short essay we give a definition characterized from a strictly economic viewpoint. Economics is based on the premise that individuals make the choices that are best choices for them. What is best depends on the resources available, on the constraints, on the incentives, but above all on their individual preferences, the reasons behind which are not investigated.
Regarding this, the analyses of T. Scitovsky (1976) on *intrinsic and extrinsic motivations* should be noted. Paradoxically, according to Scitovsky (1976) unlike extrinsic motivations, which are reinforced by monetary incentives, these reduce rather than stimulate engagement in activities that require strong intrinsic motivations, such as those of volunteering, because the work in this case is stimulating in itself and the addition of a further stimulus (the incentive) contributes to weaken them. Galbraith too, in "The new Industrial State - The General Theory of Motivation" (1979), states that: "The major error of economics has always been that everything is attributed to pecuniary innovation, economic reward".

<div align="center">***************</div>

From here on, throughout the paper, the two protagonists of the exchange are named "A" (buyer) and "B" (seller) respectively.

As an example, we can refer to a classic situation: that of a tourist/buyer at a stall in an Eastern market where price haggling is the rule.

The seller sees that the attention of the tourist has settled on an object the price of which is not shown. To determine the " negotiation dance " , as H. Raiffa[3] calls it, there are, at the start, the ***reserve prices*** of the two subjects, i.e. **the *maximum price that the tourist is willing to pay*** and ***the minimum price at which the seller is willing to sell***. Each subject does not know the reserve price of the other, but both formulate a hypothesis about it and from that position, starts the negotiation that, step-by-step, leads to the agreement.

At the beginning, therefore, both the tourist and the seller evaluate the possible **ΔU** that concerns them (underline: profitability).

If, for example, the tourist were a collector who recognized an object that is missing from his collection, he would be very motivated to purchase it and therefore his reserve price, which he carefully hides, would be quite high. If the seller for his part was in a difficult economic situation, he would also be very motivated to close the deal and therefore his reserve price could be quite low, close to the cost he incurred to obtain the item.

In any case, the tourist (A - buyer) will have a motivation (**M$_A$**= **ΔU$_A$**) to which an initial reserve price will correspond $p_{RA}$ and similarly the seller (B) will have his motivation (**M$_B$**= **ΔU$_B$**) to which the initial reserve price will be linked $p_{RB}$.

Each subject, that is, in each negotiation, even *before assessing* the probable motivation of the other, taking into account in some cases also external references, hypothesizes his reserve price: in the case of the buyer the maximum price he is willing to spend ($p_{RA}$); in that of the seller, the minimum price at which he is willing to sell ($p_{RB}$).

When each individual makes an assessment of the other's probable reason for the exchange, the reserve prices identified by each may be altered.

In particular: if the buyer (A) understands that the seller (B) is strongly motivated to sell (called from here on **M$_{BA}$**= $\Delta U_{BA}$ the motivation of B for his own gain, as perceived by A), then he will tend to lower his reserve price. Similarly, if the seller understands that the buyer is strongly motivated to buy (similarly called **M$_{AB}$**= $\Delta U_{AB}$ the motivation of A perceived by **B** for the good / service which B can supply) then he will tend to raise his reserve price. It is believed that, in the ***perceptions*** of the individual contracting parties, even albeit vaguely, there are relationships between these variables (**M$_A$ M$_B$ M$_{AB}$ M$_{BA}$**) that can concretely alter the starting reserve prices of each player, bringing them to different $p_{RA}^* \ p_{RB}^*$ values that are affected by each person's assessment of the motivation of the other party.

The element that goes to alter the first estimate of the reserve price by the two subjects ($p_{RA}$, $p_{RB}$) is the **imbalance perceived** between the two motivations: if in fact both sides perceived a strong motivation by the other against an equally strong personal motivation, the original reserve prices would not be affected. If instead one of the two, being mildly motivated, feels that the other is instead strongly motivated to the exchange, then he would alter his reserve price to obtain an advantage: all this naturally <u>before</u> the negotiation has started.

A possible relation, to support an analysis on the topic, could be (distinguishing the evaluations of the two subjects) the following:

$$p_{RA}^* = p_{RA} \cdot \frac{M_A}{M_{BA}} = p_{RA} \cdot \frac{\Delta U_A}{\Delta U_{BA}} \Rightarrow p_{RA}^* \leq p_{RA} \ if \ M_{BA} = \Delta U_{BA} \geq M_A = \Delta U_A$$

2)

where all the elements are evaluated by the Tourist (A - buyer)

$$p_{RB}^* = p_{RB} \cdot \frac{M_{AB}}{M_B} = p_{RB} \cdot \frac{\Delta U_{AB}}{\Delta U_B} \Rightarrow p_{RB}^* \geq p_{RB} \ if \ M_{AB} = \Delta U_{AB} \geq M_B = \Delta U_B$$

3)

where all the elements are evaluated by the Seller (B).

The two formulas represent pretty much what occurs in the <u>initial</u> moments of an exchange process: looking at the first, which expresses the tourist's perceptions, if he realizes that the seller is strongly motivated to sell (**$M_{BA}$** high), he will lower his reserve price. So the seller, if he perceives a strong motivation from the tourist (**$M_{AB}$** high), would raise his reserve price.

If both perceived, even if in different ways, a sort of parity in the reasons for the exchange, the $\frac{M_A}{M_{BA}} \ or \ \frac{M_{AB}}{M_B}$ ratio would be irrelevant and would not change the initial assessments.

In this case, the $p$ value with which the negotiation will be settled will be included (generally positioned halfway) between the two reserve prices:

$$p_{RA} > p > p_{RB}$$

If one were to evaluate it in an interview in the field, the $\frac{M_A}{M_{BA}} \ or \ \frac{M_{AB}}{M_B}$ ratio should be verified as a **"perception of imbalance"** between the two motivations, as it is experienced by each of them, and certainly not as a relationship between two single elements that are difficult to assess.

## 2) The element of Power.

However, the heart of this analysis is based on the certainty that in a huge number of cases that arise in everyday experience it is not enough to introduce the imbalance in motivations. We must bring into play an element of extraordinary importance: the **imbalance of power**.

However, to arrive at the imbalance of power, it is first of all necessary to give a definition of power, *limited to the sphere of exchange*, which makes it possible to identify the possible forms of imbalance.

To stress that this is a complex operation we believe it is appropriate to recall a reflection by Steven Lukes[4]:

*"…In daily life and in scholarly works, we discuss its (power's) location and its extent, who has more and who less, how to gain, resist, seize, harness, secure, tame, share, spread, distribute, equalize or maximize it, how to render it more effective and how to limit or avoid its effects. And yet, among those who have reflected on the matter, there is no agreement about to how to define it, how to conceive it, how to study it and, if it can be measured, how to measure it…"*.

So defining power is a gamble. In the following, however, we limit ourselves to proposing a definition that can be applied to a rather narrow field: that of the exchange between two people.

***Within these limits, therefore, we define the "power" of a subject, within an exchange, as his (<u>perceived</u>) ability to produce a significant variation of well-being, in a positive or negative sense, for the subject with whom he interacts, relinquishing well-being in his turn to produce it[5]***

Having adopted this definition, we call A and B the two subjects involved, and correspondingly $K_A$ and $K_B$ the power of A on B and of B on A:

$K_A = \Delta U_{BA} - \Delta U_{AA}$   (referring to the good / service which A can supply to B)          4)

$K_B = \Delta U_{AB} - \Delta U_{BB}$   (referring to the good / service which B can supply to A)          5)

Where $\Delta U_{AA}$ and $\Delta U_{BB}$ is the own well-being cost, by A and B, to create the change of well-being in B by A ($\Delta U_{BA}$) and in A by B ($\Delta U_{AB}$).

***According to this definition <u>the stronger the disproportion</u> between the perceived benefit or damage that a subject can provide to the other and the cost in terms of well-being that he has to bear, <u>the higher</u> his power is.***

It should be borne in mind, however, that the power of A perceived by A ($K_A$) while negotiating with B, on the basis of the good / service that <u>A can supply to B</u>, is as one would predict, different from the power of A perceived by B, for the same good / service. Similarly, the power of B perceived by B ($K_B$) while negotiating with A, on the basis of the good / service that, in the exchange, <u>B can supply to A</u>, is different from the power of B perceived by A.

---

[4] Lukes, Steven, 2005 (Second Edition). *"Power: a Radical View"*. Palgrave Macmillan Houndmills, Basingstoke, Hampshire RG21 6XS and 175 Fifth Avenue, New York, N.Y. 10010. 60-69

[5] The definition proposed is not far off that of Galbraith (1983,1986) who identified three "instruments" of power which he called, respectively, **condign power** (the power to inflict unpleasant or painful alternatives), **compensatory power** (the ability to reward) and **conditional power** (the ability to change beliefs through education and persuasion). This definition is to be found in : Galbraith, John Kenneth, 1983, *"The Anatomy of Power"*, published in the United States of America by Houghton Mifflin Harcourt, Boston.

We must take into account what we have said by calling **K$_A$** the perception that A has of his /her own power and **K$_{AB}$** the power of A as perceived by B. Similarly, we proceed by calling **K$_B$** the perception that B has of his/her own power and **K$_{BA}$** the power of B as perceived by A.

To back this up, we can provide some examples of how power is perceived in the ordinary sense: someone who, with a single phone call, obtains a job for another person is considered to have strong power; a mafia boss who gives the nod so that the shopkeeper who does not pay the *protection money* has his shop burned down, is believed to have a strong power. In both cases, a small effort on the part of those in power corresponds to a great benefit or a great damage to the subject to whom the "powerful person" relates to. In some ways *the welfare functions of two different subjects* are compared.

Similarly to the case in which the only component at stake is the "motivation", when each subject makes an assessment of the "power" that the other can exert over him in terms of ability to affect his own well-being, the reserve prices ($p_{RA}$, $p_{RB}$) roughly identified by each may be altered.

In particular: if the buyer (A) for example understands that the seller (B) does not have the capacity to exercise a significant form of power over him (**K$_{BA}$** low) then he will tend to lower his reserve price. If on the other hand the seller understands that the buyer has an obvious ability to affect his well-being (**K$_{AB}$** high) then he will also tend to lower his reserve price.

A relation is then proposed, between the elements at stake, which is an extension of the one previously formulated between a buyer and a seller:

$$\left( p_{RA}^* = p_{RA} \cdot \frac{M_A}{M_{BA}} \, \frac{K_{BA}}{K_A} \right) \quad \text{all elements are evaluated by A (buyer)} \qquad 6)$$

$$\left( p_{RB}^* = p_{RB} \cdot \frac{M_{AB}}{M_B} \, \frac{K_B}{K_{AB}} \right) \quad \text{all elements are evaluated by B (seller)} \qquad 7)$$

The buyer and seller make *different estimates* of the two relationships, but both reason in similar ways.

For the purposes of a field survey, the **M$_A$/M$_{BA}$ - M$_{AB}$/M$_B$** and **K$_{BA}$/K$_A$ - K$_B$/K$_{AB}$** ratios must both correspond to the *imbalances perceived* by each subject: imbalances in motivation and power, imbalances that actually increase or decrease reserve prices. Beyond the formal representation adopted, there is no doubt that, as a result of the mental calculation that each person implements in unknown ways, the original reserve prices are eventually changed.

****************

In the following, we consider situations that seem to be market–based, but which are not so in reality, as one of the two subjects involved has no alternatives.

Let us consider some cases in which the imbalances of motivation and power are clearly present, in the sense defined in the formulas 6) - 7).

A type of relationship is that established, for example, between a *gang master* and a laborer to work in the fields. We will call the gang master "A" (Buyer) and the laborer, "B" (Seller).

- The reserve price of the gang master $p_{RA}$ (hourly) is, at the beginning, roughly, close to the average current price for that type of work[6]. The gang master may not be strongly motivated to recruit the laborer if he has many others available: his $M_A$ (buyer) would be low in this case. If from various indications he understands that the laborer (seller) is in serious difficulty due to problems of how to survive ($M_{BA}$ high) he would take it into account when making his offer. Refusing to employ him, the gang master has the power to cause serious problems to the laborer: therefore his $K_A$ is high while $K_B$ is obviously almost nil. At the limit, on the basis of the formula, $p_{RA}^*$ could reach indecent values, in the order of a fraction of € or $ per hour, enough for pure survival. It is an event that occurs with great frequency in similar situations.

The proposed formula, as a pure reference framework, seems to reflect quite well what happens in terms of perceiving the opportunities and the constraints on the part of both subjects.

Let us now consider another case in which similar forms of imbalance occur in the evaluations that precede the path, be it long or short, towards agreement.

This is the relationship established between an employer in a company and a worker over 50 years old, who has lost his job during the recent crisis. As usual we call "A" the employer (typically represented by the Human Resources manager in charge of hiring) and "B", the person who is hoping to be hired.

- The reservation price $p_{RA}$ (or rather the monthly salary that the employer is able to offer) is, as in the previous case, fairly well defined by the market, for the type of job that can be done by B (seller). On the basis, however, of indications gleaned in the interview, the employer understands that the motivation of B is very high due to family problems ($M_{BA}$ high). He is also aware of having other applicants with equally valid CVs ($M_A$ low). Therefore, he has the power to choose someone else, causing serious damage to B in a period when there are few job offers: i.e. $K_A$ is high and at the same time $K_B$ is low. The employer is inclined to lower his $p_{RA}^*$ .

- For his part, B is strongly motivated and defends his position as well as he can, but understands the situation: although waiting for the moment of bargaining, if there is one, but already knowing that he will have to accept what is offered, which is certainly below, and perhaps a lot below, the average that was his reference point.

As a third case, we consider a fairly complex situation: an exchange relationship characterized by corruption.

Consider the relationship established between a building contractor (A) with few scruples who needs a license in a hurry, and a council official of the relevant government office (B) who is in charge of the case.

Also in this case we can speak of reserve prices both for the entrepreneur and for the government official, considering the bribe prices paid in similar situations that depend on the gravity of the infringement and the value of the land to which the license refers. In this case, however, compared to the previous ones, the motivation of the entrepreneur to have the license and the official to have the bribe money do not present special imbalances. As for the possibility of A to damage B ($K_A$) and B to damage A ($K_B$) it is presumed not to

---

[6]

In reality, the definition of hourly pay regards the group of laborers that the gang master recruits and transports, but the elements that come into play in the *gang master* -laborer relationship are those described above making the hypothesis that the relationship involves an individual.

present strong imbalances as both subjects are prosecutable for corruption. Neither of the two would report the other. The starting reserve price of both in this case would therefore not be affected.

- If for some reason, however, there was a strong gap between the offer of the entrepreneur and the request of the official, if the entrepreneur was a powerful person with political support, almost immediately, in the face of veiled threats, the official would perceive the *imbalance of power* and would be forced to lower his demands. In this case the reserve prices would be affected, falling for both subjects, in probably different ways that leave some room for negotiation.

With this example, you can refer to all those cases (and there are many) in which there is a **controller** who lets himself be bribed by a **controlled**: all cases in which a public official has the power to report an infringement of the rules or to grant a permit. A typical example is the financial police, but also the police where they are fighting drug dealers and the mafia. There is always a price that can be paid to remove an obstacle, and power imbalances affect the outcome of bargaining a lot.

### 3) The negotiation

Attention is now turned to the <u>*initial moments of a negotiation*</u>, those in which the adjustment of the reserve prices occurs.
At this point, one begins what H. Raiffa calls, as we have already said, *the negotiation dance*: it is a complex process that involves various strategies that make use of hazards, fictions, concealments, emphases and so on. It develops in steps and is therefore difficult to represent. At the beginning ***each tries to understand the reserve price of the other*** and makes an offer ($X^B_0$ $X^A_0$ - Fig. 1) that he would like to be close to it.

We now represent, in a simplified form, a process that takes place over time, hinting only in Figure 1 at the steps that would gradually represent the successive offers but then resorting, on the right of the figure, for a more fluid representation, to the progress of functions (**$X^A$; $X^B$**) which, starting from the values of the **offers** ( $X^B_0$ $X^A_0$ ) of the seller and the buyer, tend asymptotically to the reserve prices of each.

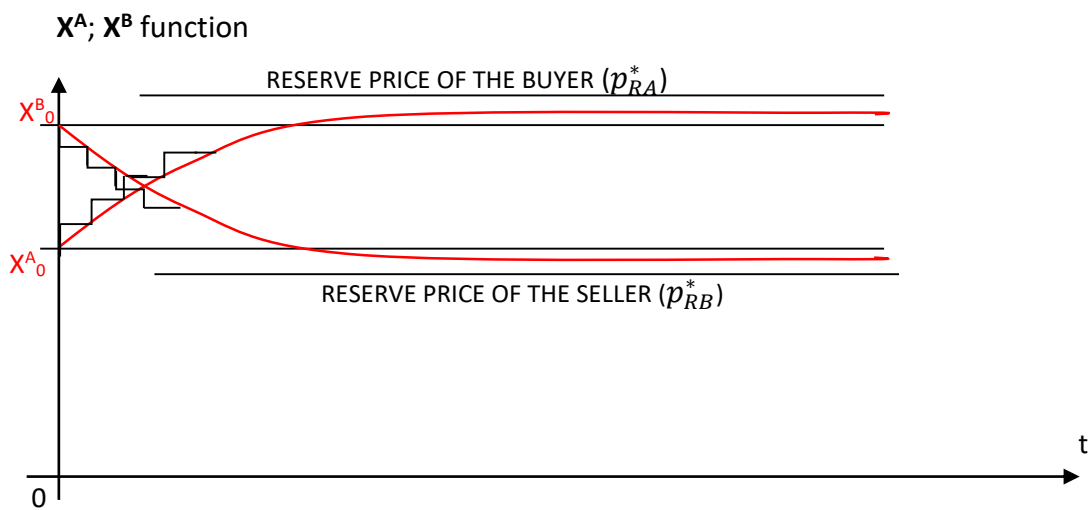

Fig. 1

To give shape to the equations of the two curves starting from $X^B_0$ and $X^A_0$ we consider a generic time interval between two negotiation phases: $t_n \rightarrow t_{n+1}$.

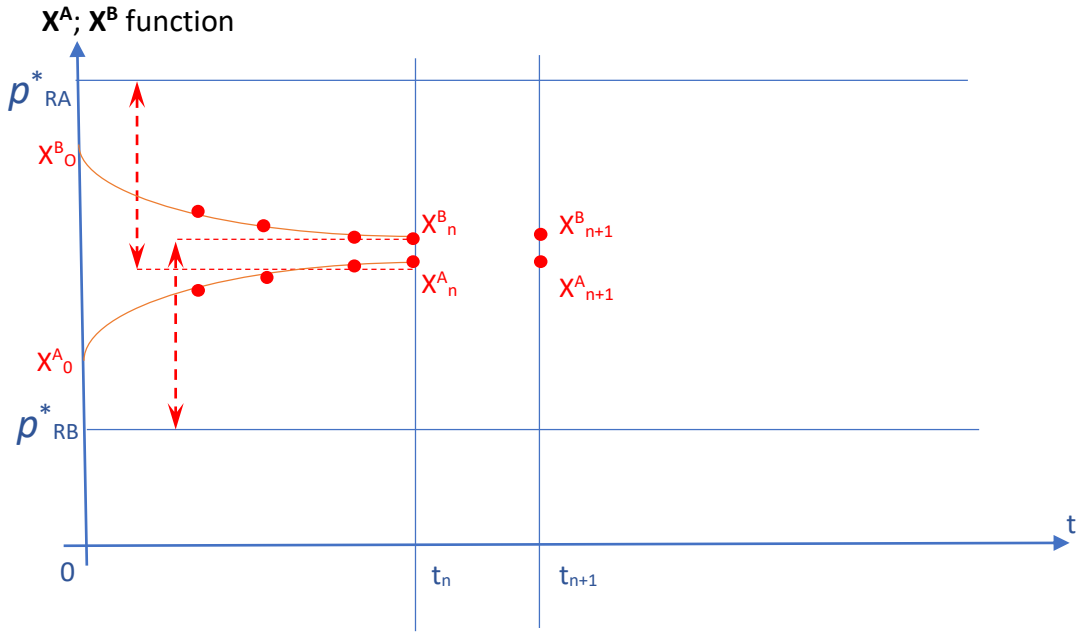

**$X^A$; $X^B$ function**

Fig. 2

Each subject at the time "n" decides what value to propose at the time "n + 1" according, on the one hand, to the distance he/she has reached from his/her reserve price; on the other hand based on the desire to slow down or speed up the closure depending on how far the proposed values are from each other, at the time "n". This type of strategy, which is hypothesized to be pursued by both parties, can be represented by the following differential equation to differences:

$$X^A_{n+1} = X^A_n + r_a(p^*_{RA} - X^A_n) + r_a'(X^B_n - X^A_n) \qquad 8)$$

$$X^B_{n+1} = X^B_n - r_b(X^B_n - p^*_{RB}) - r_b'(X^B_n - X^A_n) \qquad 9)$$

The buyer adds two increments whose value depends on the $r_a$ and $r_a'$ coefficients. The first is related to the propensity to yield; the second to the will, or not, to close the deal.

The seller, with different coefficients, behaves similarly but gives something up. The values of the "r" chosen by A and B reflect their strategies.
If it concerns a strong subject and a weak subject, obviously the former will tend to give up very little, while the latter will tend to yield a lot at every step for fear of not reaching a conclusion.
In general, when there are imbalances of motivation and power between the contracting parties, the $r_A$ $r_A'$ $r_B$ $r_B'$ coefficients are affected by the intensity of the perceived imbalances: that is, they are functions of linear approximation of the ratios

$$\frac{M_A}{M_{BA}} \frac{K_{BA}}{K_A} \text{ and } \frac{M_{AB}}{M_B} \frac{K_B}{K_{AB}} \text{ perceived respectively by A and by B.}$$

*Power and motivation therefore affect not only reserve prices, but also the negotiation process.*

To simulate what really takes place in the negotiation, we could introduce "time" as a variable that allows us to represent the possible variations in behavior of the two parties *during* the process, but it would result in formal complications that we do not wish to tackle here.

Working with Wolfram-Mathematica on the two equations above and then giving values to the parameters "r", to the initial offers and to the reserve prices, it is possible to derive the trends of the curves that characterize the negotiation.

The following is the aspect that the two curves assume in a specific case in which there is a strong buyer who does not give in, and a weak seller who suffers.

Assuming $X^B_0$= 4,5 €/h; $X^A_0$= 2,5 €/h; $r_A$= 0,05; $r_A^{'}$= 0,02; $r_B$= 0,3; $r_B^{'}$= 0,2; $p^*_{RA}$ = 5€/h; $p^*_{RB}$=2 €/h the result is as follows, when shown on the graph.

**$X^A$; $X^B$ function**

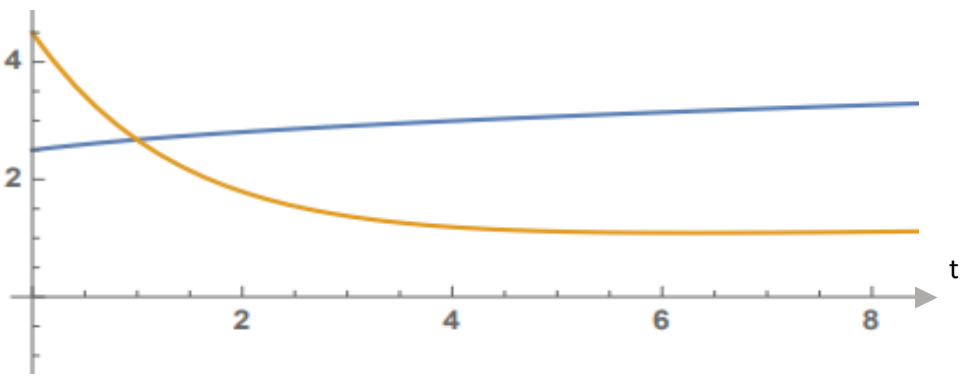

Fig. 3

The price then reached in the negotiation is very close to the initial offer of the Buyer. It can therefore be concluded that the imbalances perceived by the two parties affect both the value of the reserve prices tacitly defined by each one at the start of the negotiation, and on the final value with which the negotiation ends.

With the reflection proposed here, through very simple formalizations to which mental processes correspond in the two subjects which are not easy to identify, we reach the conclusions that have always been known to all: a very motivated subject, for example for survival problems, accepts without opposition the conditions that are imposed by a subject who has the power to get him out of a pitiful situation.

Relations 6) and 7) can be interesting for experimentation in the field, as they bring into play *imbalances perceived* by each subject that can be translated into *coefficients greater than or less than one,* which in terms of wishes, tricks, and fears or dreams of each person, can concretely and significantly alter the final result of an exchange with respect to what would happen in "normal" situations: those in which the relationships of motivation and power are comparable.

*****************

**4) The supply chains.**

This analysis can be extended to *chains of exchanges* such as those that occur in *supply chains*. Today these chains, thanks to the low transport costs for large quantities, can cross different continents, such as those that starting (backwards) from European countries end up in Vietnam or Korea or Bangladesh.

In the following, as an example to start with, typical of a certain place, we present in brief the characteristics of the tomato supply chain, which in Italy is divided into two districts: the northern district (Parma, Piacenza) and the southern district (Foggia, Potenza). In particular, the southern district has structural anomalies that cause strong tensions between the parties involved. [7]

The production is marked by the presence of something that has already been mentioned in this paper: the "caporalato", which is anyway going to disappear in the medium term with the progressive intoduction of automated collection. The corporals recruit labor, generally from Africa: they provide housing, food and transport to the fields: it is they who establish the pay rate. Mostly it is piecework.

The farmers refer to the Producer Organizations (PO) that play a mainly bureaucratic role as they register the contracts and oversee the procurement of European funds: they are the go-betweens to the Processing Industries that process and package the product.

The latter mainly deal with the Large-Scale Distribution which establishes the prices for products supplied by companies with online auctions carried out in two stages (the second part starting from the lowest price of the first auction).

The more fragmented the production area, the weaker the Producers Organization is in its relationship with the companies, and therefore the lower the harvest prices, which can squeeze workers' wages to levels that at most allow survival. The maximum bargaining power is in the hands of large retailers.

We can see two phases in the negotiation process that involves the various players in the supply chain: in the first, agreements are established based on market forecasts and production capacities of the producers. In the second phase the final prices are negotiated at various levels according to the actual demand and the quality / quantity of the product available.

In both phases there is, on the one hand, the power of the large-scale retail trade which, having a fairly clear idea of the selling prices that can be accepted by consumers, pushes down the prices to be paid to companies in order to increase their profits. On the other hand, as a consequence, companies force down prices from producers, especially if the producers' organization (PO) is weak. Finally, the price squeeze falls on workers who, not enjoying any trade union protection, are forced to accept survival wages.

It is not a sequence of normal market exchanges, but one in which the power of each player is decisive: it is a chain of relationships between the strong and the weak. In some cases, the producer with some expedient tries to increase his power: for example, by initially declaring a high production capacity, but then sowing less than required. When harvesting, he / she announces smaller quantities, causing an increase in prices.

---

[7] La crisi dell'industria del pomodoro tra sfruttamento e insostenibilità – a cura di Fabio Ciconte e Stefano Liberti TERZO RAPPORTO della campagna #Filiera Sporca - 2016 (The crisis of the tomato industry- between exploitation and unsustainability - by Fabio Ciconte and Stefano Liberti THIRD REPORT of the campaign #Filiera Sporca - 2016)

But usually the more we move towards the roots of the supply chain, the more we reduce the bargaining power of those involved. The greatest imbalances of power occur in the last stages of the chain, where the production of the raw materials is to be found.

Below, we analyze two extreme cases, starting at the roots of the respective supply chain.

The first case refers to Brazil:

- If from the foundry products made in Brazil we go back to the coal used to produce them and to some of the areas where the coal is mined (for example the "batterias" of Mato Grosso do Sul) we discover situations characterized by very strong imbalances of power between employers and workers. In those areas there are people called "Gatos" who have the task of recruiting workers for the batterias in the favelas or in any case in the poor areas close to Mato Grosso.
- During the recruitment the Gato promises well-paid work, food and the opportunity to visit families with a certain frequency. The Gato enjoys great power as he promises good work and is fully aware of the high motivation of the people he is recruiting. Recruiting is easy.
- Once the workers have been transported to the batterias, everything changes. The place is inside the forest and far from population centers, the workers must hand over their identity cards and their workcards, and they are already in debt for the costs of transport. Work in the furnaces is at the limit of human endurance.
- Consequently: the power of the Gato is very high, the power of the workers null, motivation of the workers very high as the alternative is death. The remuneration of workers is zero in most of the batterias, as confirmed by formulas 6) and 7) above, where A is the Gato and B the individual worker.

The second case refers to Pakistan:

-  If we go back from the world of construction to the brick makers, for a part of the production we get to Pakistan where there are (as of a few years ago) about seven thousand kilns: huge constructions in which the bricks are fired. They are prepared by households starting from the earth which is transformed into mud with water and then shaped in the moulds.
- The hierarchy of the kilns is made up of the "Munshi", the manager who responds directly to the owner and is surrounded by specialists (stackers, stokers, emptiers) and "Jamadar": the team leaders on whom the families who prepare the raw bricks depend. Often families are made up of former peasants left without land. They are mostly recruited by the Jamadar, who provide them with a roof and the essentials for living but leaving them in debt. Household income covers essential expenses: any unforeseen event (a funeral, drought, illness) involves indebtedness that must be repaid with days or months of labour.
- The work is piecework: at least a thousand bricks must be produced per family per day. Drops in productivity are not tolerated: households that do not make enough are sold, with their debt, to furnaces located in remote areas where working conditions are even worse. They are transported by truck under the control of armed guards.
- Young women and widows are subject to sexual harassment by bosses.

Returning then to the reflections from which we started, the motivation of these workers ($M_B$) is high because their survival is at stake. The power of the employer ($K_A$) is very high and that of the worker ($K_B$) is null. It follows that the initial expectation of income of the worker, also linked in this case to the promises in the recruitment phase, is totally wiped out: the money that the workers receive is just enough for survival and the debt never goes down.

At the root of these chains there is slavery.

This happens in the case of gold mines, diamond mines, tobacco, carpets, clothes: the list is long and even involves trafficking in human organs. These are extreme cases: but it is often true that *the latter have few alternatives* and are forced to lower the price of their product to the extreme limit where the gain is cancelled.

Then the fluctuations of the market that occur in the upper part of the supply chain are *reversed backwards* due to the greater or lesser contractual power of the subjects that trade at the various stages.

Taking into account the formation of the prices discussed above, both for exchanges involving only two subjects and for chains that collect multiple subjects in sequence, it is considered possible, going from the micro to the macro, to construct a model that represents the functioning in economic terms of a community characterized by power imbalances at all levels.

<p align="center">***************</p>

### 5) The exchange of goods or services.

Until now we have considered the process that leads to the identification of the ***price*** that eventually the buyer (such as the employer) will pay to the seller (the worker).

In the following we will consider those exchanges in which there is **no money** at stake and there is **no negotiation**, generally speaking: there is an agreement or a refusal.

Without the mediation of money, we enter an area in which the focus of attention is the well-being of the subjects who are directly affected: everyone makes a **balance** between what they can lose in well-being by providing their own good / service and how much they can earn by accepting the good / service offered by the other person.

In order to support the reflection, a picture is presented in figure 4, in which the left part is dedicated to the subject **A** and the right part to **B**.

On the left side we can distinguish: above the negative variation of utility (in red) that the subject A undergoes when he gives the good/service to B and, below, the utility variation, positive, which benefits A when he receives the good/service offered by B. In violet the *initial level* of well-being of A.

In completely theoretical terms the final level of well-being to which each could find himself after the exchange would be obtainable as an algebraic sum of what was lost and of what was gained, starting from the initial level of well-being.

On the right side, the same variations appear in reverse order for B.

**They are symbolic representations of *<u>perceived values</u>* that allow each subject to evaluate the convenience to make the exchange**.

As shown in the figure, a vertical reading allows you to highlight the gap between what each one loses and what he gains: what we defined at the beginning as ***motivation.***

Horizontally, instead, we see on the upper right side the impact 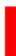 that the good / service offered by A to B, represented in the central part of the figure with the symbol , has on the welfare of B ($\Delta U_{BA}$ if you imagine it as perceived by A). Given that, the sale of the asset has a negative impact on the well-being of A ($\Delta U_{AA}$), it follows that horizontally we can highlight, ideally, $K_A = \Delta U_{BA} - \Delta U_{AA}$ which we have defined as the power of A perceived by A. If we think of the same symbols as perceived by B, the figure could show $K_{AB}$.

Similarly the lower part, representing the sale of another good/service from B to A, allows the evaluation of **K<sub>B</sub>** or **K<sub>BA</sub>.**

The meaning of the gray areas will be explained later, but in general terms they represent external elements which could modify the original term of trades.

Fig. 4: A symbolic representation of welfare changes to which "A" and "B" are subject, due to the exchange of goods / services from A to B and from B to A.

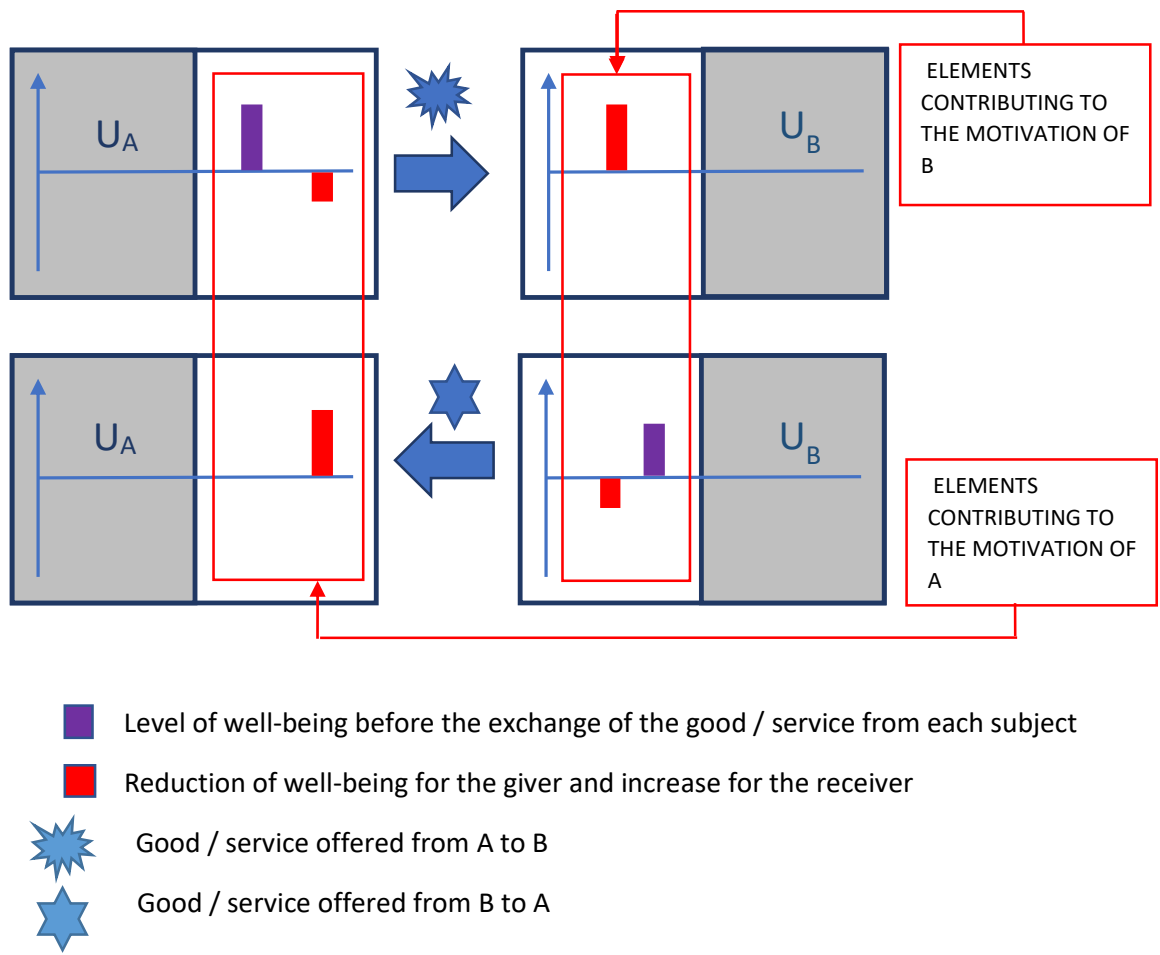

■ Level of well-being before the exchange of the good / service from each subject

■ Reduction of well-being for the giver and increase for the receiver

✸ Good / service offered from A to B

✶ Good / service offered from B to A

*****************

Assuming that A is a **strong subject** and B is a **weaker person**, the sale of the asset offered by A (upper area of the fig. 4) will determine a small variation in well-being in A and a large variation in well-being in B, while the sale of the asset on the part of B (lower area of the figure) could have a very high cost: to obtain A's well-being it may be that B has to greatly decrease his own.

Looking vertically, the strong subject does well from the exchange, while the weak one, starts at a disadvantage, and in order to have an advantage from the exchange and return to the positive, he must accept that he should provide the product or service requested by A, suffering a large decrease in well-being.

Fig 5: Symbolic representation of welfare changes to which "A" and "B" are subject due to the exchange of goods / services from A to B and from B to A, where A is strong and B is weak.

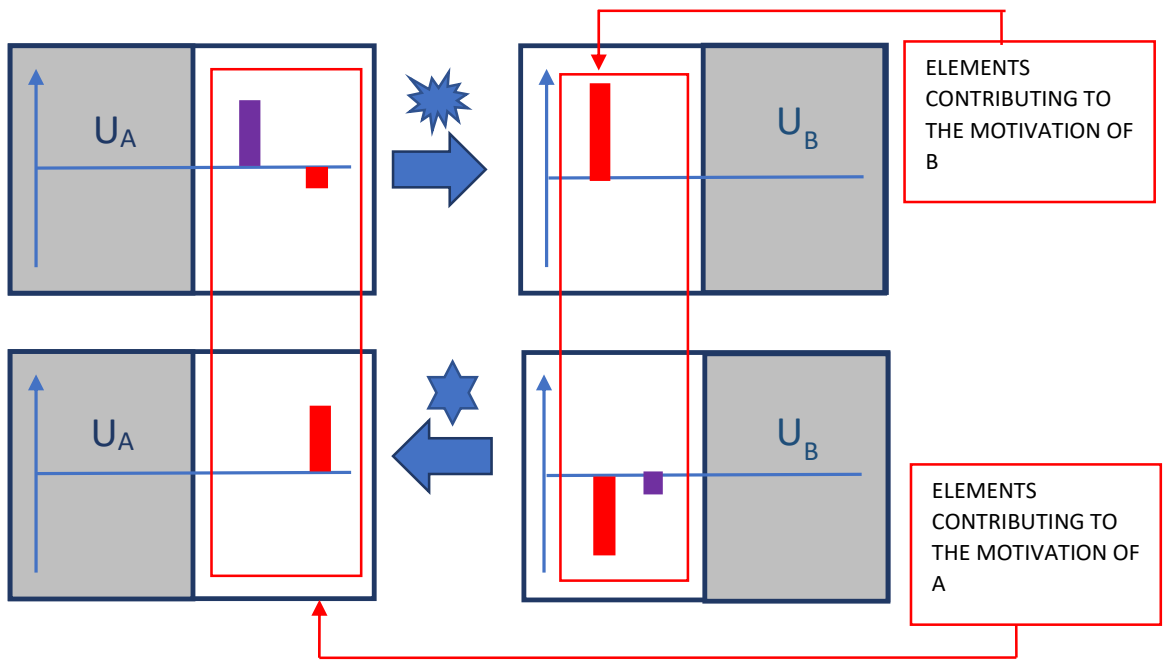

■ Level of well-being before the exchange of the good / service from each subject

■ Reduction of well-being for the giver and increase for the receiver

✦ Good / service offered from A to B

★ Good / service offered from B to A

*****************

The **gray areas** located on the left for A and on the right for B are spaces in which specific situations may be shown, where external elements influence, even greatly, the exchange **without being connected to the goods at stake**. These may be threats that involve potential changes in well-being or external context elements that can reinforce or weaken the power of A or B respectively.

: Symbolic representation of welfare changes to which "A" and "B" are subject due to the exchange of goods / services from A to B and from B to A and <u>external elements</u> that alter the welfare balance by introducing effects of potential actions (e.g. threats) unrelated to the exchange.

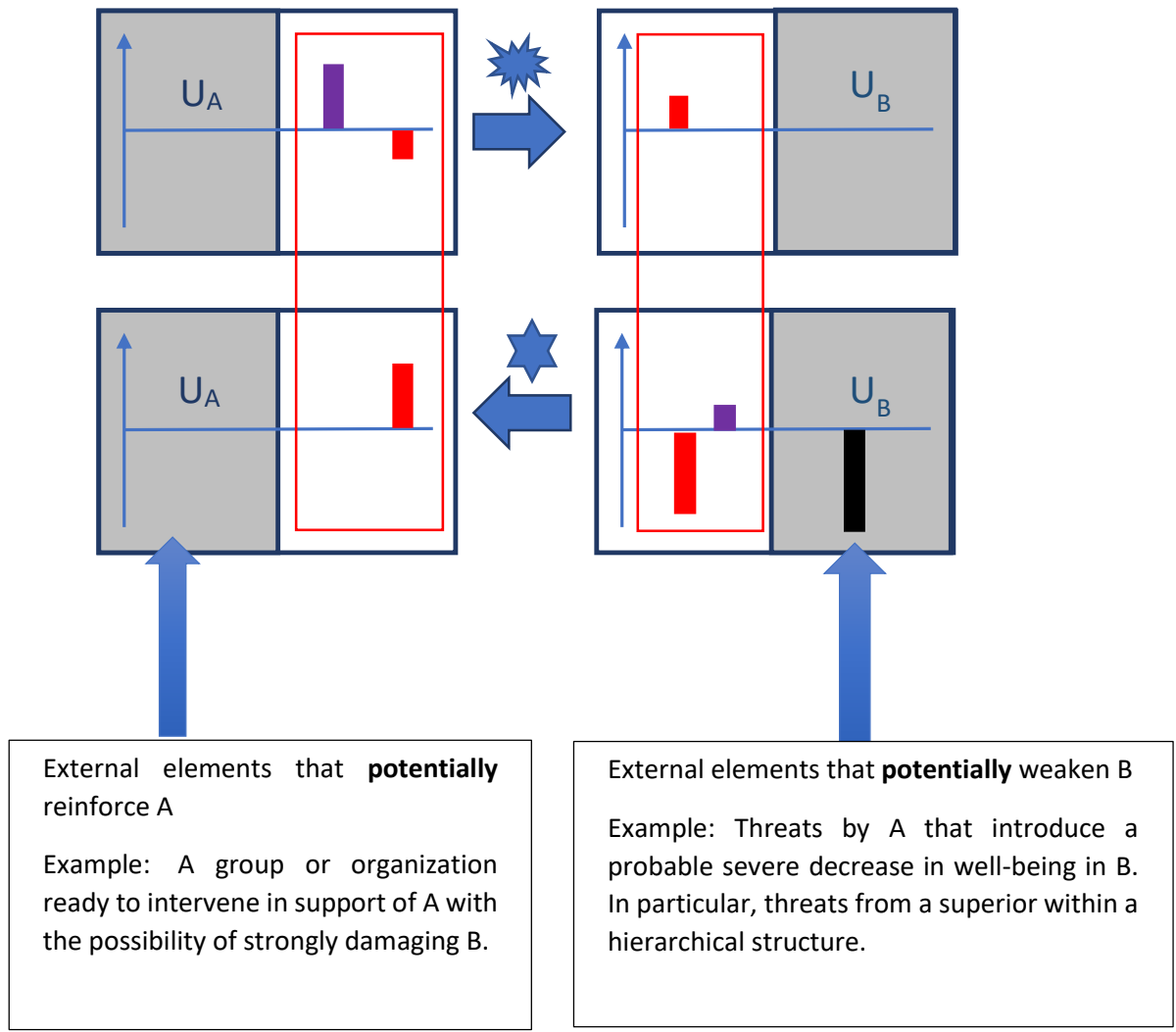

External elements that **potentially** reinforce A

Example: A group or organization ready to intervene in support of A with the possibility of strongly damaging B.

External elements that **potentially** weaken B

Example: Threats by A that introduce a probable severe decrease in well-being in B. In particular, threats from a superior within a hierarchical structure.

**As a whole, the framework provides a symbolic representation of *the circles that are activated in an exchange.***

As an example, which is referred to in the table under fig. 6: if A threatens B, B <u>has to accept</u> the conditions of exchange set by A, which may already be negative for B, <u>in order not to end up in an even more negative balance</u>: the threat can be represented with a negative (potential) change in well-being in the gray area (side B).

This category includes relations that are established in a hierarchy, for example within companies. The subordinate (B) may have little motivation to carry out what the superior (A) requires, but the relationship of power, to avoid future consequences, makes him do what is required of him.

As seen in the table at fig.6, A might have the backing of an organization that supports him and that constitutes a potential threat to B.

For his part, B, in cases different from that shown, can be protected by an environment made up of laws, cultures and institutions that shield him from the threats: all this can be translated case-by-case into a ***potential*** variation of well-being that, if activated, reinforces or weakens the power of each subject.

<center>***************</center>

The question that arises is: *under which conditions* is exchange accepted and *with what power imbalance*?

Reading the figure 4 vertically, if $M_A$ and $M_B$ are both positive, then the exchange is accepted. But if a threat on B looms, the exchange can be accepted even if $M_B$ is negative.

If, however, in the gray areas, elements that reinforce or weaken the power of A or B appear, ***indirect power*** must be taken into consideration; a potential variation of well-being extraneous to the object of exchange. This form of power alters the terms of the balance and makes the weak party consent to the exchange, because the refusal would lead to worse conditions.

The $K_B / K_A$ ratio, again in theory, provides an imbalance indicator. As to how the subjects then carry out these evaluations, through their mental processes, it is difficult to understand: the power **imbalance** is always perceived, as a whole and not through its elements.

<center>***************</center>

We would like to make some further reflections with reference to the hypothesis that A is a strong subject and B is a weak subject. We will review two types of exchange, of the type we are considering, with the sole purpose of highlighting how power plays a role in determining various forms of imbalance including violence.

**First type**: the strong subject A "promises" an asset and asks for a *countervalue*: the weaker B assesses whether it suits him, also trying to estimate the probability that the promise is maintained (if A does not keep his promise, B would not able to react because he has little power).

- This type is present in cases where a "selection" is being made: selection of extras or secondary actors to make a film or a play, a selection of models for fashion shows, a selection of shop assistants for temporary contracts in shops or supermarkets, a selection of exhibitors for trade fairs, a selection of secretaries. The list could be long.
- In these cases, the selector is the strong subject (*strong in that situation*) who can promise the candidate gets the job in return for sexual favours.
- If the candidate is strongly motivated because he/she needs money, it may be that he/she considers it worth acquiescing: in terms of utility functions, the increase in well-being linked to the overcoming of primary needs in order to get by, at least for a while, *may be higher than* the discomfort associated with providing unwanted sexual favours. Therefore, if the initial situation is of great need, the motivation to accept the conditions set by the selector is likely to be high. This is frequently the case.

**Second type**: the strong subject promises a benefit but at the same time threatens the weak subject with punitive actions if he does not accept the conditions imposed in exchange for the benefit. It is the case in which the weak subject has *no alternatives* and must submit to the will of the strong. It is a completely imbalanced exchange in which the motivation is provided by the strong subject, *with an element unrelated to the exchange.*

- The classic case is that of the "protection money" demanded of a shopkeeper by someone working for a mafia type organization (the gangster is a strong subject *because* he has a mafia organization behind him). The shopkeeper does not ask for protection: he is offered it if he is willing to pay the protection money, but at the same time it is made clear that, if he does not accept the conditions, his shop might be in danger and so might his family too. The police in the area are corrupt and do not protect shopkeepers.
- The power of the strong person is also high because he is protected from the outside. The motivation of the weak is high. It is impossible to reject the conditions offered without suffering very serious consequences. The environment behind the weak person does not protect him. He must therefore accept the conditions even if, with respect to the starting welfare situation, his motivation for the exchange is negative. This is a very widespread situation in all the areas controlled by the mafias.

There are many cases that fall into these categories, but all fall into the "welfare/discomfort circle" that we have shown in Fig. 4.

As we have said, exchanges of this type that do not include prices to be paid cannot be used for the construction of models that represent the functioning on the economic level of a community that, for example, presents widespread power imbalances. However, the scheme proposed in Fig.4 may constitute the logical basis for the construction of an indicator that measures the **"*perception of equity*"** in an exchange.

For example, the following indicator could be used to perform this function, consisting of dimensionless ratios, averaged with respect to <u>subjective evaluations</u>:

$$I = \frac{M_A}{M_B} \frac{K_B}{K_A}$$

10)

The indicator is 1 if there are no imbalances. If instead A (strong subject) has a lot of power and normal motivation, while B (weak subject) has little power and strong motivation, the indicator lowers and the exchange is classifiable as not fair. If B is supported and gains power while A is constrained, for example, by institutional constraints that protect the weak, then the index rises indicating that the exchange is fairer. This is just an example.

**Equity indicators** of this type can allow us, through field sampling, to "**map**" the equity in the different areas and strata of a territory or a community.

Returning to figure 4 and thus to the cases in which there is *external support* for the exchange:

-If the <u>strong</u> is a bully, he will have people behind his group waiting to intervene should the boss call.

- If the <u>strong</u> is the general who attacks a young recruit, he will have a system ready to hide the facts.

- If the <u>strong</u> is a pedophile priest, he will have behind him a powerful organization that, at least in some cases, protects him.

We prefer to consider separately the case of the <u>weak</u>, in the reflections that follow.

<p style="text-align:center">***************</p>

### 6) Defence strategies:

It should be noted that the weak (as well as the strong), in the context of these considerations is limited to the issue of exchanges, and not a *particular category* of people. The <u>*context*</u> and the <u>*object of exchanges*</u> are the elements that determine the level of weakness of a subject.

A powerful professional who finds himself with a broken-down car and a dead mobile phone at three o'clock at night in a suburban street of a big city is in a weak state: he will have to ask for help from the first person he meets, but at that hour he may fall foul of bad characters. He will have to face an exchange situation without defenses of any kind: it can end badly. The probability of meeting a "strong" person because he has a knife in his pocket (a strong person at that moment) is high: he could, without effort, take away everything, including the car, if he manages to get it to work. And the weak person, would not be able to resist.

We try to say in the following, with some examples which will question the types of power, how a <u>weak person</u> can be defended by an environment and how he can defend himself with individual strategies.

a) How a weak person can be protected by a <u>favorable context</u>:

- In general terms, it can be said that there are categories that are chronically weak when situations change: the poor and the disabled. An evolved civil society takes on the task of defending this type of weak person. It does so through the laws, the institutions responsible for respecting them, values and culture that spreads within the media and the schools. What results then is that acts of violence against defenseless people find an environment which is immediately hostile, ready to react to protect them. This does not happen in authoritarian and corrupt regimes where the agents of the institutions make personal use of their power.

- A weaker person can receive protection by belonging to a group that highlights its problems with the support of the media: in this case, for example, the unpaid workers who go on the roof of the warehouses or cranes, the demonstrations in the town squares or in front of the city halls of the earthquake victims to whom economic aid for reconstruction has not arrived, the torchlight processions against the local underworld that kills those who hinder its plans.

b) How a weak person can defend himself on an <u>individual basis</u>:

In Steven Lukes' already quoted text[8] are some statements by James Scott[9] including the following:" ... the powerless are often obliged to adopt a strategic pose in the presence of the powerful ".

---

The prevailing strategic attitude is presented as that of those who, while having within themselves a feeling of anger and rebellion against injustice and the abuses that they are forced to suffer, must lie, smile even, and hold on.

But there are also other ways to go *in order to extend the definition of power to the ability to get help from one subject to another with whom one has a history of agreements, projects and collaborations or a friendship or emotional relationships*.

If we add these kinds of people to the weak and powerful we have talked about so far, then we can think of putting in place *sequences of power relationships* that can also give power to a weak person as an end result.

We try to clarify this with two examples: the first from literature, the second a real case.

> In "The Purloined Letter" by Edgar Alan Poe we can see a defense strategy of a blackmailable character (the weak) which is rather anomalous and interesting: a perfidious and intelligent minister (the strong), who in meeting a person of rank, belonging to the royal family, manages to steal a letter, proof of a secret love affair, with which he realizes that he can blackmail him together with the lady involved. The victim of the theft knows who stole the letter and asks the Prefect of Police, his acquaintance and trusted friend, to recover it without publicizing the fact, promising him a very high reward. The Prefect uses all his skills to find the letter, by searching the minister's house, in his absence, with all the most sophisticated means available to the police, but he cannot find it. Discouraged, the Prefect, offering a reward, asks for help from his friend Dupin, who knowing how cunning the minister is, goes to see him and discovers that the letter is actually in plain sight on a card holder, but with the envelope upside down, on which the seal of the minister appears on the outside instead of that of the royal family which remains on the inside. Using a trick, he removes it and replaces it with a similar one, but with a message inside addressed to the minister who he hates, for having committed wrongs in the past. The minister does not know it but his power to blackmail has vanished and the lady involved in the affair is safe. The sequence is: victim-Prefect, where there is an economic power together with a relationship of trust; Prefect-Dupin where there is the economic power, but above all Dupin's previous hatred towards the blackmailer and the honor of the lady. Dupin solves the case not with police methods, like the Prefect, but by following the devious mechanisms of the blackmailer himself.

> - If an employee of a company that is not suffering financial problems is induced to resign because he is suddenly offered low-profile roles at a political moment when the union has no strength and the labor market does not offer opportunities that require the type of skills that that employee has, then the employee can only ask for a soft landing. But he has no contractual strength. He then resorts to a close relative who has a friend who is director of the staff of a large company, asking for help. The friend commits herself: she has daily business contact with a law firm that supports the company exactly in the making of layoffs, with which she does a lot of high-value business. In this case she asks, as a favor, to the lawyer with whom she has a close relationship, to play the opposite role with the company that is forcing out the person she is (indirectly) protecting. The lawyer willingly accepts and attends the redundancy negotiation without asking to be paid for his services. His law firm is authoritative and renowned for its achievements. In the negotiation, the parties agree for a more than satisfactory redundancy payout.

These kinds of chains, which can be called **Power Chains**, are actually used with great frequency everywhere, even if often unconsciously.

***They are based on the fact that a person who is weak in one way may be strong in another, and in different ways.***

This is a mechanism that requires imagination: the pathways, characterized by the fact that in each step the subject that intervenes is stronger than the next, are not immediately visible: we must build them abandoning the logic of common sense and the mental habits of our daily life. But it is an important tool that *can make a weak person strong*.

To close with the defense strategies, it should be remembered that today network technologies can help us along unimaginable roads: if a child chased by a pedophile could secretly press a button that simultaneously alerted parents, siblings, if there are any, and the nearest police station (we have all the technology needed and it is even cheap) the pedophile could easily be stopped and arrested. Likewise, for those women who, when working in shifts, have to return to deserted streets late, it would be crucial to have a dedicated tool that would allow them to alert the nearest police station. It would be a much more effective defense system than having the patrols that police cars normally have at set times.

**7) From local to global**:

Taking into account what has been said so far, namely the importance of the power dimension in simple exchanges and supply chains, it would be very interesting to build two *virtual models* of societies, with the analytical tools we currently have, where:

- In one, there prevails an authoritarian and corrupt structure that determines at all levels the presence of imbalanced exchanges and supply chains in which at different stages, for the determination of prices, all the power of the strongest person is used.
- In the second one, there are institutions and strong union organizations able, with the control of laws and contracts, to guarantee the balance in the various forms of exchange that take place at different levels of the social structure.

It would be interesting with simulations to understand what effects the two different ways of operating the institutions have on the economic level: what effects they can have on the *development potential of society* and on the *differences between wealth and poverty*.

This is certainly a complex simulation, but would have, if possible, an extraordinary training potential in economics courses. I realize that in this thought there is a dreamlike component that has too much weight, but perhaps with the tools available today it is not entirely unfeasible.

However, it is certain that the well-being and malaise of the various subjects operating in the two societies cannot be summed up: we know, however, that in the first model, well-being remains concentrated in the upper layers of society while the malaise is the dominant element in the lower layers of the population. We know that the gap between wealth and poverty is particularly strong in countries that adopt the authoritarian model.

But we must add that, if we move away from the ideal model envisaged in the second type of society, in which all exchanges are brought back to balanced relations thanks to the intervention of institutions and, we look at what actually happens in Western democracies, then the imbalances return in the various exchanges at various levels and in the supply chains; imbalances that then lead to polarization between wealth and poverty.

In any case we are convinced that ***not taking into account*** in the economic analysis[10] the element of power, present at all levels of real society, does not lead anywhere if the characteristics of the world being examined are those highlighted in Fig. 7.

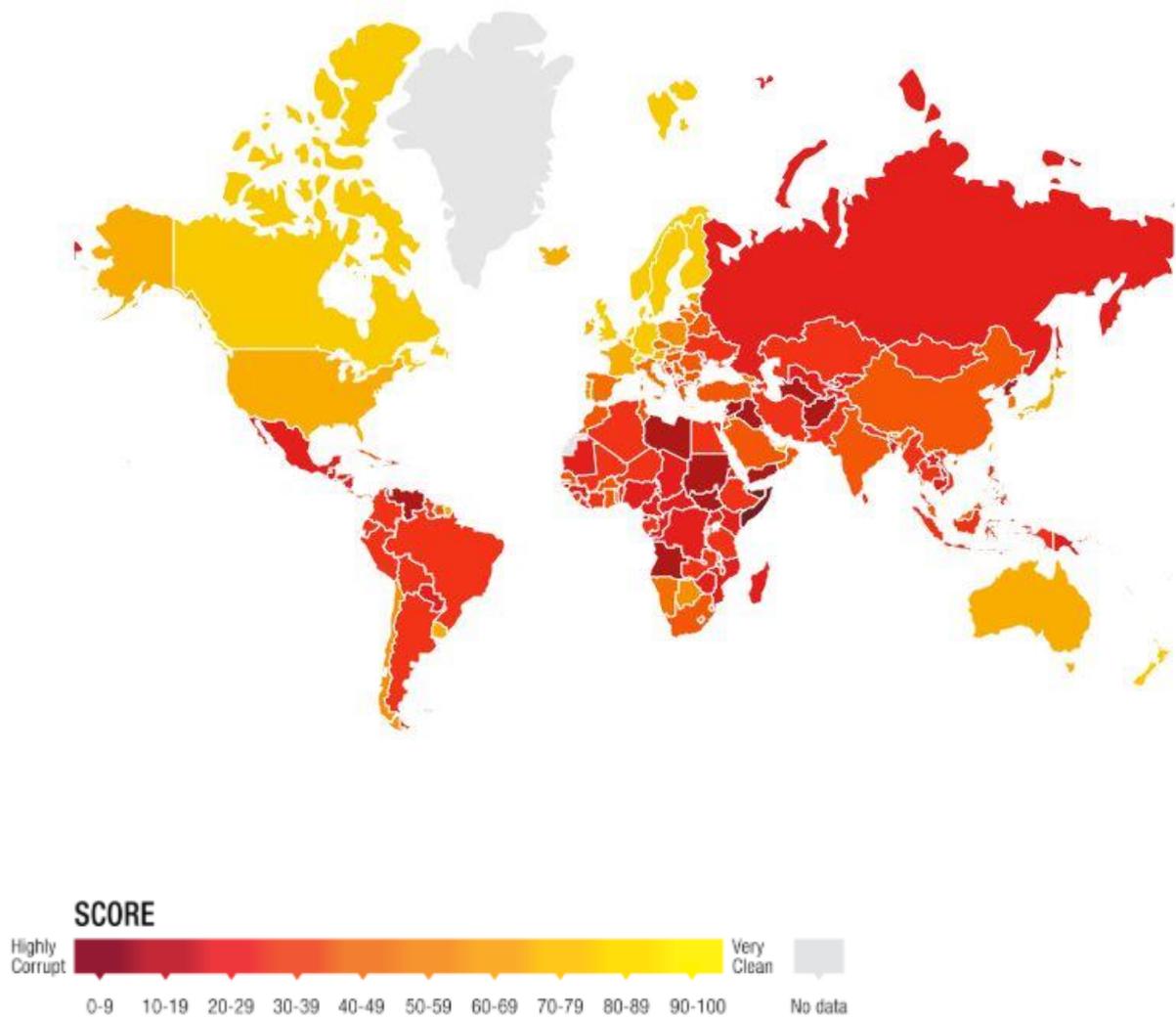

Fig. 7. Corruption Perception Index 2018 - Source: Transparency International.

---

[10] The following statement is significant in Galbraith (1973): "The decisive weakness in neoclassical economics…is not the error in the assumptions by which it elides the problem of power…Rather in eliding power – in making economics a non-political subject – neoclassical theory, by the process, destroys its relation with the real world. J. K. Galbraith (1973). "Power and the Useful Economist", *American Economic Review*. V. LXIII, 1, (March).

**Bibliography**:

The following list contains only the texts that provided crucial points for the drafting of the Paper.

 N. B.: The reflections summarized in this paper come from far away and are the result of a long study carried out in seminars held on the subject at the Department of Economics of the University of Trieste which involved professors of statistics, economics and psychology. I must confess that, as often happens, the ideas that have been developed in the various discussions, have *suddenly* found a place, I believe coherent and presented here, to which indeed the texts quoted above have contributed significantly.

***************